\documentclass[12pt]{article}
\usepackage{latexsym}
\usepackage{amssymb}
\usepackage{epsfig}
\begin{document}
\title{ Dynamical origin of low-mass fermions in Randall-Sundrum background}
\date{}
\maketitle
\begin{center}

Kenji Fukazawa$^{\small 1}$, Tomohiro Inagaki$^{\small 2}$, 
Yasuhiko Katsuki$^{\small 3}$, Taizo Muta$^{\small 4}$ and 
Kensaku Ohkura$^{\small 5}$\\

{\it \small{
$^{\small 1}$ 
Department of Mechanical Engineering, Kure National College of
Technology,\\ Kure 737-8506, Japan \\
$^{\small 2}$ Information Media Center, Hiroshima University,  
Higashi-Hiroshima \\ 739-8521, Japan \\
$^{\small 3}$ Faculty of Social Information Science, Kure University, 
Kure, Hiroshima 737-0182, Japan\\ 
$^{\small 4}$ Hiroshima University, Higashi-Hiroshima 739-8511, Japan \\
$^{\small 5}$ Research Center for Nanodevices and Systems, 
Hiroshima University, Higashi-Hiroshima 739-8527, Japan  \\
}}

\end{center}

\begin{abstract}
We investigate a dynamical mechanism to generate fermion mass 
in the Randall-Sundrum 
background. We consider four-fermion interaction models where the fermion field propagates 
in an extra-dimension, i.e. the bulk four-fermion interaction model. It is assumed that two 
types of fermions with opposite parity exist in the bulk.
We show that electroweak-scale mass is dynamically generated for a 
specific fermion anti-fermion condensation, 
even if all the scale parameters in the Lagrangian
are set to the Planck scale.
\end{abstract}
\clearpage
%%%%%%%%%%%%%%%%%%%%%%
\section{Introduction}
%%%%%%%%%%%%%%%%%%%%%%
The Standard Model (SM) provides a remarkably successful description
of known phenomena. On the other hand the SM has an unsatisfactory
feature which is called hierarchy problem, disparity between the 
Planck scale and the electroweak scale. 
One of the solution for the hierarchy problem is found in the 
higher dimensional theory \cite{arkani}. In the scenario the SM scale can 
be obtained from the ratio between the Planck scale and the size 
of the extra dimension. 

Randall and Sundrum proposed an alternative approach to solve
the hierarchy problem in a five-dimensional curved spacetime \cite{randall}.
They considered the five-dimensional anti-de Sitter 
spacetime compactified on an orbifold, $S^1/Z_{2}$, and two 3-branes 
existing at the orbifold fixed points. The 3-branes are four-dimensional 
subspaces embedded in the five-dimensional spacetime. It was shown
that the spacetime metric satisfies the Einstein equation and the electroweak scale 
$O$(TeV) is derived from the Planck scale $M_{P}$ without introducing 
any very large parameters. 

In the beginning it is considered that the SM particles are constrained on 
the 3-brane.  But there is a possibility the SM particles also propagate in 
an extra dimension, i.e. bulk SM. Goldberger and  Wise pointed out that 
the bulk scalar fields have modes whose mass terms are exponentially 
suppressed on a brane as well as the brane particles \cite{goldberger}. 
We can identify the lightest modes of bulk fields to be the SM particles. There 
are a lot of possibilities to put the SM particles in the bulk \cite{chang, hewett}. 
In the Randall-Sundrum background some mechanisms
are proposed to generate an extremely light fermion mass
like a neutrino \cite{grossman}.

It is the standard scenario that an elementary Higgs boson induces
mass of particles through Higgs mechanism. 
The dynamical mass generation, for example top quark condensation \cite{miransky} 
in extra dimensions \cite{dobrescu}, is an appealing alternative scenario. 
It has been known that the spacetime curvature plays an important
role to dynamical symmetry breaking \cite{ina}.
A four-fermion interaction model is studied in the RS background
as a prototype model of dynamical symmetry breaking \cite{abe}.
The dynamical origin to generate the low mass fermion is
found by assuming the existence of the strong interaction
between the bulk fermion and the brane fermion.
Gauge theories is also considered in the RS background by using
the Schwinger-Dyson equation \cite{inagaki}. It is pointed out that the 
strong interaction is naturally appears in the gauge theory.
It is found that the SM scale is obtained on the brane from 
the bulk QCD coupled with the brane fermion.

Dynamical origin of the low mass fermion is discussed
in some models with brane fermions. In the present paper 
we study the possibility to generate the low fermion mass 
dynamically starting from the theory only with the bulk 
fermion. Assuming the four-fermion like interaction
between the bulk fermions, we evaluate the phase structure 
and the natural mass scale of the model.
This paper is organized as follows: In section 2 we explain our 
setup and model.
In section 3 we analyze the effective potential and fermion mass spectrums. 
In section 4 we make a comment on the solution for the hierarchy problem in our model.
In section 5 we give the summary and discussions.

%%%%%%%%%%%%%%%%%%%%%%%%%%%%
\section{Bulk Four-Fermion Model}
%%%%%%%%%%%%%%%%%%%%%%%%%%%%
As is known, the chiral symmetry prohibits
a Dirac mass term in four-dimensional spacetime. 
The Dirac mass term is generated through breaking of the chiral symmetry.
Especially in a four-fermion model a composite operator of fermion and
anti-fermion, $\bar{\psi}\psi$,
may develop a non-vanishing vacuum expectation value and the chiral symmetry is broken dynamically \cite{Nam}.
Here we study a bulk four-fermion model in the RS spacetime. The RS spacetime is the five-dimensional spacetime 
which is compactified on an orbifold $S^{1}/Z_{2}$ of radius $r_c$.

The metric of the RS spacetime is described by
\[
 G_{AB} = diag (e^{-2 k|y|} \eta_{\mu \nu}, -1 ),
\]
where y is the coordinate of an extra-dimensional direction. 
There is no chiral symmetry in the RS spacetime.
To see it we consider a free bulk fermion theory, 
\begin{equation}
S_{F}= \int d^4x dy \sqrt{G} \left[ \bar{\psi} i\Gamma^{\bar{A}}e_{\bar{A}}^{A}
(\partial_{A}+ \frac{1}{8} \omega_{A}^{\bar{B}\bar{C}}[\Gamma_{\bar{B}},\Gamma_{\bar{C}}]
) \psi \right],
\label{action1}
\end{equation} 
\\
where $e_{\bar{A}}^{A}$ is the inverse of the vierbein and 
$\omega_{A}^{\bar{B}\bar{C}}$ is the spin connection. 
We denote the five-dimensional Dirac $\gamma$ matrix by
$\Gamma_{\bar{A}}=(\gamma_{\mu},i \gamma_5)$.
The action is invariant under $y \rightarrow y\pm 2\pi r_c$, 
and therefore we can restrict y to $-\pi r_c < y \le \pi r_c$. 
The action is also invariant under $y \leftrightarrow -y$.
The fermion field, $\psi$, has even or odd property under 
five-dimensional parity transformation,

\begin{eqnarray}
\left\{
\begin{array}{l}
\psi(y)\rightarrow \gamma_{5} \psi(-y) = \psi(y) ; even,   \\
\psi(y)\rightarrow \gamma_{5} \psi(-y) = - \psi(y) ; odd.  \\
\end{array}
\label{parityap}
\right.
\end{eqnarray}
For even-parity fermion five-dimensional spinor fields can be 
expanded in terms of Kaluza-Klein(K-K) modes,
\begin{eqnarray*}
\psi(x,y)&=& \psi_{R}(x,y)+ \psi_{L}(x,y) \nonumber \\
&=& \sum_{n=0}^{\infty} \psi_{R}^{(n)}(x)g_{R}^{(n)}(y)
+\psi_{L}^{(n)}(x) g_{L}^{(n)}(y). \label{kkap}
\end{eqnarray*}
The parity transformation (\ref{parityap}) for even case gives  
\begin{eqnarray}
\left\{
\begin{array}{l}
g_{R}^{(n)}(y) = g_{R}^{(n)}(-y), \\
g_{L}^{(n)}(y) = -g_{L}^{(n)}(-y). \\
\end{array}
\right.
\label{kkbcap}
\end{eqnarray}
From the periodicity on y and  
the latter equation of (\ref{kkbcap}) 
we can easily see $g_{L}^{(n)}(0) = g_{L}^{(n)}(\pi r_c) = 0$.
For the bases which diagonalize the Lagrangian 
in terms of the K-K modes, the action (\ref{action1}) reads
\[
S_{F}= \int d^4x \bar{\psi}_{R}^{(0)} i \partial_{\mu} \gamma^{\mu}
 \psi_{R}^{(0)} + 
\sum_{n=1}^{\infty} \bar{\psi}^{(n)} (i \partial_{\mu} \gamma^{\mu}-m_n)
 \psi^{(n)} ,
\]
where $\psi^{(n)}=\psi_{R}^{(n)}+\psi_{L}^{(n)}$.
The boundary condition $g_{L}^{(n)}(\pi r_{c})=0$ yields 
$ m_n= k \pi n / (e^{k \pi r_{c}}-1)$ \cite{chang}.
The properly normalized mode functions are given as follows: 
\begin{eqnarray}
\left\{
\begin{array}{l}
g_{R}^{(0)}= \sqrt{\frac{k}{1-e^{-k \pi r_c}}} e^{-\frac{1}{2}k \pi r_c}
e^{\frac{1}{2} k|y|}, \ \ g_{L}^{(0)} = 0, \\
g_{R}^{(n)}= \sqrt{\frac{2k}{1-e^{-k \pi r_c}}} e^{-\frac{1}{2}k \pi r_c}
e^{\frac{1}{2} k|y|} \ (n \geq 1), \\ 
g_{L}^{(n)}= \sqrt{\frac{2k}{1-e^{-k \pi r_c}}} e^{-\frac{1}{2}k \pi r_c}
e^{\frac{1}{2} k|y|} \ (n \geq 1).  \\
\end{array}
\right.
\label{rskkap}
\end{eqnarray}

Since $g_{L}^{(0)}=0$, only the right handed zero mode survives and
other modes are vector-like.
When the five-dimensional parity is odd,
the spinor field is expanded as
\begin{eqnarray}
\psi(x,y)&=& \psi_{R}(x,y)+ \psi_{L}(x,y) \nonumber \\
&=& \sum_{n=0}^{\infty} \psi_{R}^{(n)}(x)g_{L}^{(n)}(y)
+\psi_{L}^{(n)}(x) g_{R}^{(n)}(y), \nonumber
\end{eqnarray}
where the mode functions $g_R$ and $g_L$ are defined in (\ref{rskkap}).

Only one of the zero modes survives for both fermions. It is always a
massless fermion because there is no chiral partners in the induced
four-dimensional spacetime. We would like to pursue the scenario 
that chiral symmetry breaking generates the mass of fermions. 
To realize the chiral symmetry in four-dimensional spacetime 
it is necessary to introduce the five-dimensional parity even fermions 
in addition to the parity odd fermions as the chiral partner. 
It is possible that the right and left handed zero modes
form the mass term via the four-fermion interaction \cite{chang}.

The induced four-dimensional Lagrangian of the bulk four-fermion model is given by
\begin{equation}
{\cal L} = \int dy \sqrt{G} [\bar{\psi}_{1} i \partial_A \Gamma^{A} \psi_{1}
+ \bar{\psi}_{2} i \partial_A \Gamma^{A} \psi_{2}
- \hat{\lambda} (\bar{\psi}_1 \psi_2)(\bar{\psi}_2 \psi_{1})],
\label{lagrangian1}
\end{equation}
where $\psi_1$ and $\psi_2$ are parity even and odd fermions respectively. 
Because the natural scale of bulk is the Planck scale, $M_p$, 
it is natural to take $\hat{\lambda}$ \ as $O(1/M_p^3)$.
This Lagrangian is invariant under a discrete chiral transformation;
\begin{eqnarray*}
\left\{
\begin{array}{lll}
\psi_{1}(x,y) &=& \gamma_5 \psi_1(x,-y), \\
\psi_{2}(x,y) &=& - \gamma_5 \psi_2(x,-y), \\
\end{array}
\right.
\end{eqnarray*}
which prohibits the mass term $m \bar{\psi_1} \psi_2$.
Introducing the auxiliary field, $\sigma \sim \bar{\psi}_1 \psi_2$, we rewrite the Lagrangian;
\begin{eqnarray}
{\cal L}
 = \int dy \sqrt{G} \left[
\left(
   \begin{array}{cc}
    \bar{\psi}_1 & \bar{\psi}_2 \\ 
   \end{array}
 \right)
\left[
   \begin{array}{cc}
      i \partial_A \Gamma^A & -\sigma^{*} \\ 
      \sigma &  i \partial_A \Gamma^A \\ 
      \end{array}
 \right]
\left(
   \begin{array}{c}
       \psi_1  \\ 
       \psi_2  \\ 
      \end{array}
 \right)
-\frac{| \sigma |^2}{\hat{\lambda}}
		     \right].  
\label{lagrangian2}
\end{eqnarray}
After applying the K-K mode expansion the Lagrangian reads
\begin{eqnarray}
{\cal L}
 &=&  
\bar{\psi}_{1R}^{(0)} i \gamma^{\mu} \partial_{\mu}
 \psi_{1R}^{(0)} +\bar{\psi}_{2L}^{(0)} i \gamma^{\mu} \partial_{\mu}
 \psi_{2L}^{(0)} \nonumber \\ \nonumber \\
&+& \hspace{-0.5cm} \sum_{1 \leq n} \left[\bar{\psi}_{1R}^{(n)} i \gamma^{\mu} \partial_{\mu}
 \psi_{1R}^{(n)} +\bar{\psi}_{1L}^{(n)} i \gamma^{\mu} \partial_{\mu}
 \psi_{1L}^{(n)} +\bar{\psi}_{2R}^{(n)} i \gamma^{\mu} \partial_{\mu}
 \psi_{2R}^{(n)} +\bar{\psi}_{2L}^{(n)} i \gamma^{\mu} \partial_{\mu}
 \psi_{2L}^{(n)}\right] \nonumber \\
&+& \hspace{-0.5cm} \sum_{0 \leq m,n} 
  \left(
   \begin{array}{cccc}
    \bar{\psi}_{1R}^{(m)} & \bar{\psi}_{2R}^{(m)} &
    \bar{\psi}_{1L}^{(m)} & \bar{\psi}_{2L}^{(m)} \\
   \end{array}
 \right)
{\Large M}
  \left(
   \begin{array}{c}
    \psi_{1R}^{(n)} \\
    \psi_{2R}^{(n)} \\
    \psi_{1L}^{(n)} \\
    \psi_{2L}^{(n)} \\  
   \end{array}
 \right)
- \int dy \sqrt{G} \left[\frac{|\sigma|^2}{\hat{\lambda}}\right] 
, \label{lagrangian3}
\end{eqnarray}
where $M$ is the fermion mass matrix which is given by \\
\begin{eqnarray*}
{\Large M} &=& 
\left(
\begin{array}{cc}
0  & M_1 \\
M_2 & 0 \\
\end{array}
\right):  \nonumber \\
M_1 &=&
\left(
   \begin{array}{cc}
      \int\! dy \sqrt{G}[ g_{R}^{(m) *} \partial_{y} g_{L}^{(n)}]  
       & - \int\! dy \sqrt{G}[ g_{R}^{(m) *} \sigma^{*} g_{R}^{(n)}]   \\ 
    - \int\! dy \sqrt{G}[ g_{L}^{(m) *} \sigma g_{L}^{(n)}] & \int\! dy \sqrt{G}[ g_{L}^{(m) *} 
     \partial_{y} g_{R}^{(n)}]   \\
   \end{array}
\right), \\
M_2 &=&
\left(
 \begin{array}{cc}
      - \int\! dy \sqrt{G}[ g_{L}^{(m) *} \partial_{y} g_{R}^{(n)}] 
     & - \int\! dy \sqrt{G}[ g_{L}^{(m) *} \sigma^{*} g_{L}^{(n)}] \\ 
    - \int\ dy \sqrt{G}[  g_{R}^{(m) *} \sigma g_{R}^{(n)} ]
     &  - \int\! dy \sqrt{G}[ g_{R}^{(m) *} \partial_{y}  
     g_{L}^{(n)}] \\
   \end{array}
\right).\\
\end{eqnarray*}
Since the RS spacetime has no translational invariance along the y direction, 
there appears the y dependence of vacuum expectation value, $\langle \sigma \rangle$. 
The vacuum expectation value $\langle \sigma \rangle$ is determined by 
observing the minimum of the induced four-dimensional effective potential. 

%%%%%%%%%%%%%%%%%%%%%%%%%%%%%%
\section{Phase Structure in RS Background}
%%%%%%%%%%%%%%%%%%%%%%%%%%%%%%
We evaluate the induced four-dimensional effective potential and 
calculate the vacuum expectation value 
$\langle \sigma \rangle$ in the leading order of 1/N expansion. 
The effective four-dimensional action is derived 
after performing the integration over all the fermion fields, 
\begin{equation}
S_{F}=\ln \det \left[i\partial_{\mu} \gamma^{\mu}
+ M[\sigma,\sigma^{*}] \right]-
\int d^4x \int dy \sqrt{G}\left[\frac{|\sigma|^2}{\hat{\lambda}}\right].
\end{equation}
To get the effective potential we set $\sigma$ independent of x 
and obtain the effective potential, 
\begin{eqnarray}
   V_{eff}[\sigma,\sigma^*]&=&-\frac{1}{16\pi^2} {\rm Tr}
\left[
\Lambda^4 \ln[1+\frac{M^2}{\Lambda^2}]
-M^4 \ln[1+\frac{\Lambda^2}{M^2}]+M^2 
\Lambda^2 \right] 
\nonumber \\ 
&+&\int dy \sqrt{G}\left[\frac{|\sigma|^2}{\hat{\lambda}}\right]
\label{epotential1}
\end{eqnarray}
with a momentum cutoff $\Lambda$, and the trace is taken over the K-K modes. 

Here we restrict ourselves to the following two possibilities, 
$\langle \sigma \rangle = v$ and 
$\langle \sigma \rangle = v e^{k |y|}$.
Such restricted functions may not minimize the effective potential.
But note that if such functions with non-vanishing $v$ minimize the 
effective potential even within the restricted functions, we can see that  
there is more stable state than the symmetric state, $\langle \sigma 
\rangle =0$. In other words the chiral symmetry is broken down.

\subsection{$\langle \sigma \rangle =v$ case}
First we assume that the vacuum expectation value 
$\langle \sigma \rangle$ is independent of y, 
$\langle \sigma \rangle =v$.
In general we cannot diagonalize the Lagrangian [\ref{lagrangian3}] and 
therefore we rely on the numerical analysis.
The effective potential is expressed such that
\[
   V_{eff}(v)=\frac{v^2}{\lambda}-\frac{1}{16\pi^2} {\rm Tr}
\left[
\Lambda^4
\ln[1+\frac{M(v)^2}{\Lambda^2}]
-M(v)^4 \ln[1+\frac{\Lambda^2}{M(v)^2}]+M(v)^2 \Lambda^2 \right]. 
\]
Here $M(v)$ is a matrix whose 
components are $M_{mn} (0 \le m,n \le 2N_{KK})$
with 
\begin{eqnarray*}
 M_{00} &=& \frac{v}{a}\ln[1+a], \\
M_{0n} &=&
 M_{n0} =  \frac{\sqrt{2}v}{a} \left[
\cos[\frac{n\pi}{a}] \left\{C_{i}(\frac{n\pi(1+a)}{a})-C_{i}(\frac{n \pi}{a})
\right\} 
\right. \\ && \left.
+
\sin[\frac{n\pi}{a}] \left\{S_{i}(\frac{n\pi(1+a)}{a})-S_{i}(\frac{n \pi}{a})
\right\}
\right] \ \ \ (1 \le n \le N_{KK}),\\ 
M_{0n} &=& M_{n0} = 0 \ \ \ (N_{KK}+1 \le n \le 2N_{KK}),\\ 
M_{mn} &=& \frac{v}{a} 
\left[
\cos[\frac{(m+n)\pi}{a}] \left\{C_{i}(\frac{(m+n)\pi (1+a)}{a})-C_{i}
(\frac{(m+n)\pi}{a})\right\} \right. \\ && \left.
+
\sin[\frac{(m+n)\pi}{a}] \left\{S_{i}(\frac{(m+n)\pi (1+a)}{a})-S_{i}
(\frac{(m+n)\pi}{a})\right\} 
\right. \\ && \left.
+
\cos[\frac{(m-n)\pi}{a}] \left\{C_{i}(\frac{(m-n)\pi (1+a)}{a})-C_{i}
(\frac{(m-n)\pi}{a})\right\} \right. \\ && \left.
+
\sin[\frac{(m-n)\pi}{a}] \left\{S_{i}(\frac{(m-n)\pi (1+a)}{a})-S_{i}
(\frac{(m-n)\pi}{a})\right\}
\right] \\
&& (1 \le m,n \le N_{KK}), \\
M_{mn} &=& M_{nm} = m_{n} \delta_{m,n} 
\ \ \ (1 \le m \le N_{KK}, N_{KK}+1 \le n \le 2 N_{KK}), \\
M_{mn} &=& \frac{v}{a} 
\left[
\cos[\frac{(m-n)\pi}{a}] \left\{C_{i}(\frac{(m-n)\pi (1+a)}{a})-C_{i}
(\frac{(m-n)\pi}{a})\right\}\right. \\ && \hspace{-2cm} \left.
+
\sin[\frac{(m-n)\pi}{a}] \left\{S_{i}(\frac{(m-n)\pi (1+a)}{a})-S_{i}
(\frac{(m-n)\pi}{a})\right\} \right. \\ && \hspace{-2cm} \left.
-
\cos[\frac{(m+n-2N_{KK})\pi}{a}] \left\{C_{i}(\frac{(m+n-2N_{KK})
\pi (1+a)}{a})-C_{i}
(\frac{(m+n-2N_{KK})\pi}{a})\right\}\right. \\ && \hspace{-2cm} \left.
-
\sin[\frac{(m+n-2N_{KK})\pi}{a}] \left\{S_{i}(\frac{(m+n-2N_{KK})\pi (1+a)}{a})-S_{i}
(\frac{(m+n-2N_{KK})\pi}{a})\right\}
\right] \\ &&
\ \ \ (N_{KK}+1 \le m,n \le 2 N_{KK}),  \\
\end{eqnarray*}
where
$a=e^{k\pi r_c}-1,  m_n = n k\pi/a$ and  
$\lambda = 4k\hat{\lambda}/(1-e^{-4k\pi r_c})$, 
and $C_i$ and $S_i$ are the cosine-integral and sine-integral function 
respectively. We take $N_{KK}$ as $a \frac{k\pi}{\Lambda}+1$ 
such that we neglect K-K modes heavier than $\Lambda$ 
in the unbroken phase.

\begin{figure}[ht]
         \begin{center}
           \includegraphics[width=11cm]{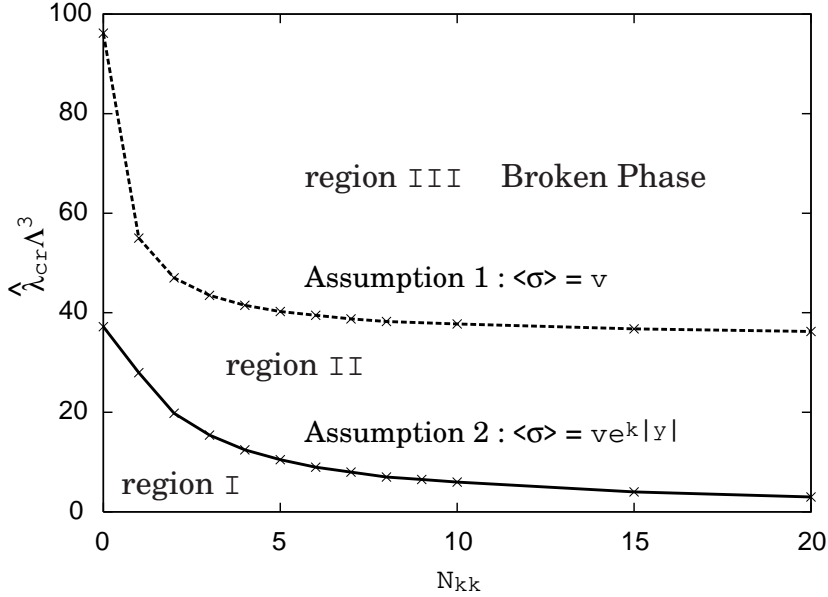}
            \end{center}
            \caption{Critical coupling constant}  
         \label{critical}
\end{figure}

For small $N_{KK}$ we can calculate the effective potential numerically. 
Through numerical inspections we  find that the chiral symmetry is
broken down at a critical coupling. 
The phase transition from the symmetric phase to the 
broken phase is of the second order.
In Fig. \ref{critical} we draw the critical coupling constant by the dotted line 
against $N_{KK}$ as $k=\Lambda=M_{P}$ fixed.

\subsection{$\langle \sigma \rangle = v e^{k |y|}$ case}
Next we consider the case, $\langle \sigma \rangle = v e^{k |y|}$. 
In this case we easily diagonalize the mass term and perform 
the y integration in Eq. (\ref{lagrangian3}).
The Lagrangian reduces to 
\begin{eqnarray}
{\cal L} &=&
   \frac{1}{\lambda} v^2
+  \left(
   \begin{array}{cc}
    \bar{\psi}_{1R}^{(0)} & \bar{\psi}_{2L}^{(0)} \\
   \end{array}
 \right)
  \left(
   \begin{array}{cc}
        i \gamma^{\mu} \partial_{\mu}   & -v \\ 
    - v  &  i \gamma^{\mu} \partial_{\mu} \\
   \end{array}
 \right)
  \left(
   \begin{array}{c}
    \psi_{1R}^{(0)} \\
    \psi_{2L}^{(0)} \\  
   \end{array}
   \right) \nonumber \\
&&\hspace{-1cm} + \sum_{n \geq 1} 
  \left(
   \begin{array}{c}
    \bar{\psi}_{1R}^{(n)} \\
    \bar{\psi}_{2R}^{(n)} \\
    \bar{\psi}_{1L}^{(n)}  \\
    \bar{\psi}_{2L}^{(n)} \\
   \end{array}
 \right)^{T}
  \left(
   \begin{array}{cc|cc}
      i \partial_{\mu} \gamma^{\mu} & 0 & m_n & -v \\
    0 & i \partial_{\mu} \gamma^{\mu} & -v & m_n \\ \hline
    m_n & - v & i \partial_{\mu} \gamma^{\mu} & 0 \\
     -v & m_n & 0 &  i \partial_{\mu} \gamma^{\mu}
   \end{array}
 \right)
  \left(
   \begin{array}{c}
    \psi_{1R}^{(n)} \\
    \psi_{2R}^{(n)} \\
    \psi_{1L}^{(n)} \\
    \psi_{2L}^{(n)} \\  
   \end{array}
 \right) , \label{lagrangian4}
\end{eqnarray}
where $\lambda$ is $2 k \hat{\lambda}/(1-e^{-2k \pi r_c})$
and $m_n=n k \pi / (e^{k \pi r_c}-1)$.
Fermion mass matrix $M$ has the following form,
\begin{eqnarray*}
M=
\left(
\begin{array}{cc}
0&
\begin{array}{ll}
m_{n}&-v\\
-v&M_n\\
\end{array}
\\
\begin{array}{ll}
m_{n}&-v\\
-v&M_n\\
\end{array}
&0\\
\end{array}
\right).
\end{eqnarray*}
The mass of the fermion is given by the eigen value of $M$, 
i.e. $v$ and $|v\pm \frac{nk\pi}{a}|$.

In the leading order of the $1/N$ expansion 
the effective potential for the Lagrangian (\ref{lagrangian4}) 
is given by 
\begin{eqnarray}
&&V_{eff} = \frac{1}{\lambda}v^2-\frac{1}{16\pi^2}\left[
\Lambda^4 \ln\left[1+\frac{v^2}{\Lambda^2} \right]
-v^4\ln\left[1+\frac{\Lambda^2}{v^2}\right]
+v^2\Lambda^2 \right] \nonumber \\
&&
\hspace{-0.5cm}
-\frac{1}{16\pi^2}\sum_{n < N_{KK}} \left[ 
\Lambda^4 \ln \left[1+\frac{v_{+}^2}{\Lambda^2}
\right] -v_{+}^4 \ln \left[1+\frac{\Lambda^2}{v_{+}^2}\right]
+v_{+}^2 \Lambda^2
\right. \nonumber \\ &&
\hspace{-0.5cm}
\left.
+ \Lambda^4 \ln \left[ 1+\frac{v_{-}^2}{\Lambda^2}
\right] -v_{-}^4 \ln\left[1+\frac{\Lambda^2}
{v_{-}^2} \right] +v_{-}^2\Lambda^2
\right],
\label{epotential}
\end{eqnarray}
where we define that $v_{+} \equiv v+\frac{nk\pi}{a}, 
v_{-} \equiv v-\frac{nk\pi}{a}$.
Differentiating the effective potential by $v^2$ we obtain the gap equation;
\begin{eqnarray*}
\frac{\partial^2}{\partial v^2}V_{eff}(v)&=&\frac{2}{\lambda}
-\frac{1}{4\pi^2}
\left[
1+\frac{2v^2}{1+v^2}-3v^2 \ln 
\left[1+\frac{1}{v^2} \right]
\right]\\  
-\frac{1}{4\pi^2}
\sum_{n \le N_{KK}} && \hspace{-1cm} \left[ 
2+\frac{2v_{+}^2}{1+v_{+}^2}-3v_{+}^2 \ln 
\left[1+\frac{1}{v_{+}^2} \right]
+\frac{2v_{-}^2}{1+v_{-}^2}-3v_{-}^2 \ln 
\left[1+\frac{1}{v_{-}^2} \right]
\right]\\
&=&0.
\end{eqnarray*}

For each numerical value of the coupling constant $\lambda$ we calculate 
the effective potential (\ref{epotential}).
It is found that the second order phase transition takes place, 
and the chiral symmetry is broken down above the critical value of 
the coupling constant.
The critical coupling constant, $\hat{\lambda}_{cr}$, is obtained 
by solving the equation;
$
[\partial^2/ \partial v^2 \ \  V_{eff}(v)]_{v=0}=0,
$
\begin{equation}
 \lambda_{cr}=8\pi^2
\left[
1+ \sum_{n \le N_{KK}} \left\{
2+ \frac{4(nk\pi)^2}{a^2+(nk\pi)^2}-6(\frac{nk\pi}{a}^2)\ln[
\frac{a^2+(nk\pi)^2}{(nk\pi)^2}
]
\right\}^{-1}
\right].
\label{criticalcoupling1}
\end{equation}

%%%%*****************************************************
%%%%*****************************************************
In Fig. 1 we show the critical coupling constant by the solid line 
against $N_{KK}$ as $k = \Lambda = M_{P}$ fixed.
In the region $II$ and $III$ the chiral symmetry is broken.
The most remarkable feature shows up in the region $II$. 
In this region the y-dependent state 
is more stable than the y-independent state.  

\section{Natural Mass Scale}

What is the natural mass scale for the lightest fermion 
in the bulk four-fermion model? 
Only a mass scale in the bulk is the Planck scale, $M_{P}$. 
We take all the mass scale in the bulk as $O(M_{P})$, 
i.e. $k\simeq \Lambda\simeq M_{P}$ 
and set $kr_c \simeq 12$, 
which gives $N_{KK}\simeq 10^{16}$. 
For such a large $N_{KK}$ the critical coupling 
(\ref{criticalcoupling1}) behaves 
to be inverse proportional to $N_{KK}$ in the 
$\langle \sigma \rangle = v e^{ky}$ case, 
\[
\hat{\lambda}_{cr} \rightarrow \frac{2\pi^2}{(1-\ln2) k^2\Lambda N_{KK}}
\simeq \frac{20}{\Lambda^3} e^{-k \pi r_c},
\] \\
which is much smaller than the natural scale $1/M_p^3$.

For $\langle \sigma \rangle = v$ case the critical coupling constant 
behaves almost like a constant value, 
$\hat{\lambda}_{cr} \Lambda \simeq O(30)$, in our numerical analysis, Fig. 1. 
It is larger than the natural scale. Therefore we conclude that the four-fermion coupling at the natural 
scale, $\hat{\lambda}\simeq\ 1/M_{P}^3$, is located in the region $II$ 
in Fig. 1. 
In this region the effective potential for $\langle \sigma \rangle = v e^{ky}$ 
is smaller than the y-independent vacuum.
If the y-dependent vacuum, $\langle \sigma \rangle \simeq v e^{ky}$, 
is a true vacuum of the theory, the mass of fermions is 
given by $v, |v \pm \pi/a|$. 
One of the fermion necessarily has mass below $\pi/a\sim O(M_{EW})$ 
independently of the value $v$.
Thus the lightest fermion mass is generated dynamically at the electroweak scale, $k\pi/a$, 
even if the vacuum expectation value $v$ is at the Planck scale $M_{P}$. 
A low mass fermion exists in the bulk four fermion model. 
It is one of the dynamical realizations of the RS mechanism \cite{inagaki, randall2}.

%%%%%%%%%%%%%%%%%%%%%%%%%%%%%%%%
\section{Summary and Discussion}
%%%%%%%%%%%%%%%%%%%%%%%%%%%%%%%%
We have investigated the bulk four-fermion model in the RS spacetime. We
assume the existence of two kinds of bulk 
fermions with different parity. It is necessary to introduce the chiral symmetry in the induced four-dimensional 
model. The effective potential is calculated in the induced model. 
Evaluating the minimum of the effective potential we found that the extra direction y-dependent vacuum is 
more stable than the y-independent one for a special region of the coupling constant. 
If we take all the mass scale in the bulk as the Planck scale, 
the lightest fermion mass is generated 
dynamically at the electroweak scale. 
Therefore a low mass fermion is obtained in our model. 
It shows the possibility to build up a realistic model which may solve the hierarchy problem dynamically in the RS spacetime. 

In the present analysis we restrict ourselves to the special form for the y-dependence of the vacuum expectation 
value. Our solution may not be the true minimum of the effective potential. To find a true minimum of the 
effective potential we calculate the effective potential (\ref{epotential1}) for a general y-dependent $\langle \sigma \rangle$.
It is interesting to calculate the stress tensor in our model and solve the Einstein equation. The y-dependent 
$\langle \sigma \rangle$ naturally change the spacetime structure. 

In the RS spacetime there are two 3-branes at the orbifold fixed point, $y=0$ and $y=\pi r$. 
The radiative correction of the brane fields has something to do with the vacuum expectation value.
But the mass scale of the $y=\pi r$ brane is the electroweak scale, $M_{EW}$. Since the influence of the 
brane fields is of the order $O(M_{EW})$, the mass scale of the lightest fermion keeps at the electroweak scale.

A SUSY extension of our model is also interesting.
We have two kinds of fermion in our model. In the RS spacetime we can
construct N=2 SUSY model, which include two kinds of fermions
automatically \cite{buchbinder}.

\section*{Acknowledgments}
%%%%%%%%%%%%%%%%%%%%%%%%%
The authors would like to thank H.~Abe and S.~D.~Odintsov for 
useful discussions.
%%%%%%%%%%%%%%%%%%%%%%%%%%%%%%%%%%%%%
%    References
%%%%%%%%%%%%%%%%%%%%%%%%%%%%%%%%%%%%%


\clearpage
\begin{thebibliography}{99}

\bibitem{arkani}
	I. Antoniadis,
	Phys. Lett. B {\bf 246}, 377 (1990); \\
	N. Arkani-Hamed, S. Dimopoulos, and G. Dvali,
	Phys. Lett. {\bf B429}, 263 (1998);
	Phys. Rev. {\bf D59}, 086004 (1999). 
\bibitem{randall}L. Randall, R. Sundrum, 
	Phys. Rev. Lett. {\bf 83}, 3370 (1999). 
\bibitem{goldberger} W. D. Goldberger and M. B. Wise, 
	Phys. Rev. Lett. {\bf 83}, 4922 (1999).
\bibitem{chang}S. Chang, J. Hisano, H. Nakano, N. Okada, M. Yamaguchi, 
	Phys. Rev. {\bf D62}, 084025 (2000).
\bibitem{hewett}
	H. Davoudiasl, J. L. Hewett, T. G. Rizzo,
	Phys. Rev. {\bf D63}, 075004 (2001). 	
\bibitem{grossman}
	Y. Grossman, M. Neubert,
	Phys. Lett. {\bf B474}, 361 (2000); \\
	N. Arkani-Hamed, S. Dimopoulos, G. R. Dvali, J. March-Russell,
	Phys. Rev. {\bf D65}, 024032 (2002).
\bibitem{miransky}
	V. A. Miransky, M. Tanabashi and K. Yamawaki,
	Phys. Lett. {\bf B221}, 177 (1989);
	Mod. Phys. Lett. {\bf A4}, 1043 (1989);\\
	C. T. Hill and E. H. Simmons, 
	Phys. Rept. {\bf 381}, 235 (2003).
\bibitem{dobrescu}
	B. A. Dobrescu, 
	Phys. Lett. {\bf B461}, 99 (1999);\\
	H. Cheng, B. A. Dobrescu and C. T. Hill, 
	Nucl. Phys. B {\bf 589}, 249 (2000);\\
	N. Arkani-Hamed, H. Cheng, B. A. Dobrescu and L. J. Hall,
	Phys. Rev. {\bf D62}, 096006 (2000);\\
	H. Abe, H. Miguchi and T. Muta, 
	Mod. Phys. Lett. A15, 445 (2000);\\
	A. B. Kobakhidze, 
	Phys. Atom. Nucl.  {\bf 64}, 941 (2001) 
	[Yad.\ Fiz.\  {\bf 64}, 1010 (2001)];\\
	M. Hashimoto, M. Tanabashi and K. Yamawaki,
	Phys. Rev. {\bf D64}, 056003 (2001);
	hep-ph/0304109;
        V. Gusynin, M. Hashimoto, M. Tanabashi and K. Yamawaki,
        Phys. Rev. {\bf D65}, 116008 (2002);
\bibitem{ina}
        T. Inagaki, T. Muta and S. D. Odintsov,
        Mod. Phys. Lett. {\bf A8} 2117 (1993);
        Prog. Theor. Phys. Suppl. {\bf 127} 93 (1997);\\
        E. Elizalde, S. D. Odintsov and Yu. I. Shilnov,
        Mod. Phys. Lett. {\bf A9}, 913 (1994);\\
        T. Inagaki, S. Mukaigawa and T. Muta,
        Phys. Rev. {\bf D52}, 4267 (1995);\\
        K. Ishikawa, T. Inagaki and T. Muta,
        Mod. Phys. Lett. {\bf A11}, 939 (1996);\\
        T. Inagaki,
        Int. J. Mod. Phys. {\bf A11} 4561 (1996).
\bibitem{abe}
	H. Abe, T. Inagaki, T. Muta, 
	in {\it Fluctuating Paths and Fields}, edited by 
	W. Janke, A. Pelster, H.-J. Schmidt, and M. Bachmann (World
	Scientific, Singapore, 2001); \\
	N. Rius, V. Sanz, 
	Phys. Rev. {\bf D64}, 075006 (2001). 
\bibitem{inagaki}
	H. Abe, T. Inagaki,
	Phys. Rev. {\bf D66}, 085001 (2002);\\ 
	H. Abe, K. Fukazawa, and T. Inagaki, 
	Prog. Theor. Phys. {\bf 107}, 1047 (2002);\\ 
	H. Abe,
	hep-ph/0307004.
\bibitem{Nam}Y. Nambu and G. Jona-Lasinio, 
	Phys. Rev. {\bf 122}, 345 (1961).
\bibitem{randall2}
	L. Randall, R. Sundrum,
	Phys. Rev. Lett. {\bf 83}, 4690 (1999);  \\
	H. Davoudiasl, J. L. Hewett, T. G. Rizzo, 
	Phys. Rev. Lett. {\bf 84}, 2080 (2000);
	Phys. Lett. {\bf B473}, 43 (2000).
\bibitem{buchbinder}
	I. L. Buchbinder, T. Inagaki and S. D. Odintsov,
	Mod. Phys. Lett. {\bf A12}, 2271 (1997);\\
	T. Inagaki, S. D. Odintsov and Y. I. Shil'nov,
	Int. J. Mod. Phys. {\bf A14}, 481 (1999).
\end{thebibliography}
\end{document}